 \documentclass[amsmath,aps,showpacs,a4paper,10pt]{revtex4}
 \usepackage{epsf}
 \usepackage{graphicx}    % Include figure files
 \usepackage{amssymb, amscd}     % AMS symbols in AMSFonts  http://www.ctan.org/pkg/amsfonts
 \usepackage{amsthm,mathenv}

\begin{document}

 \newcommand{\be}[1]{\begin{equation}\label{#1}}
 \newcommand{\ee}{\end{equation}}
 \newcommand{\bea}{\begin{eqnarray}}
 \newcommand{\eea}{\end{eqnarray}}
 \def\disp{\displaystyle}

 \def\gsim{ \lower .75ex \hbox{$\sim$} \llap{\raise .27ex \hbox{$>$}} }
 \def\lsim{ \lower .75ex \hbox{$\sim$} \llap{\raise .27ex \hbox{$<$}} }

 \begin{titlepage}

 \begin{flushright}
 arXiv:1705.06819
 \end{flushright}

 \title{\Large \bf Stability of the Einstein static universe in Eddington-inspired Born-Infeld theory}

 \author{Shou-Long~Li\,}
 \thanks{\,email address:\ sllee$_{-}\hspace{-0.275mm}$phys@bit.edu.cn}
 \affiliation{School of Physics, Beijing Institute of Technology, Beijing 100081, China}

 \author{Hao~Wei\,}
 \email[\,Corresponding author;\ email address:\ ]{haowei@bit.edu.cn}
 \affiliation{School of Physics, Beijing Institute of Technology, Beijing 100081, China}

 \begin{abstract}\vspace{1cm}
 \centerline{\bf ABSTRACT}\vspace{2mm}
By considering the realization of the emergent universe scenario in Eddington-inspired Born-Infeld (EiBI) theory, we study the stability of the Einstein static universe filled with perfect fluid in EiBI theory against both the homogeneous and inhomogeneous scalar perturbations in this work. We find that in both the spatially flat and closed cases, the emergent universe scenario is no longer viable, since the Einstein static universe cannot be stable against both the homogeneous and inhomogeneous scalar perturbations simultaneously. However, the emergent universe scenario survives in the spatially open case, while the Einstein static universe can be stable under some conditions.
 \end{abstract}

% https://www.aip.org/publishing/pacs/pacs-2010-regular-edition

 \maketitle

 \end{titlepage}

 \renewcommand{\baselinestretch}{1.0}

%============================= section 1 ===================================

\section{Introduction}\label{sec1}
An intriguing gravitational theory of connection developed by Eddington~\cite{eddington} in 1924 has drawn a lot of attention. The characteristic of Eddington's theory that the fundamental field is the affine connection at least in free de Sitter space gives a new perspective on gravity. However, this theory is incomplete since it does not consider matter. Recently, Ba\~nados and Ferreira~\cite{Banados:2010ix} proposed a new theory called Eddington-inspired Born-Infeld (EiBI) theory by coupling Eddington's theory to matter. They allow the metric to be written as a Born-Infeld-like structure~\cite{Deser:1998rj} and treat the metric and connection as independent fields rather than insisting on the purely affine action. EiBI theory is equivalent to general relativity~(GR) in vacuum. However, it has many new intriguing properties when the matter fields are included.

In past years, EiBI theory had been extensively studied in cosmology~\cite{EscamillaRivera:2012vz, Scargill:2012kg, Avelino:2012ue, Pani:2012qd, Yang:2013hsa, Du:2014jka, Cho:2014jta, Cho:2014xaa, Potapov:2014iva}, black holes~\cite{Olmo:2013gqa,  Wei:2014dka, Sotani:2015ewa}, astrophysics~\cite{Pani:2011mg, Avelino:2012ge, Sotani:2014xoa} and some related subjects (e.g.~\cite{Liu:2012rc, Shaikh:2015oha, Chen:2015eha, Arroja:2016ffm}). We refer to e.g.~\cite{Banados:2010ix} and references therein for comprehensive introduction of EiBI theory. Here we focus on one of the most attractive properties of EiBI theory, namely the big bang singularity in homogeneous and isotropic universe can be avoided reasonably. As is shown in~\cite{Banados:2010ix}, there are two ways to avoid the big bang singularity by assuming radiation domination in EiBI theory. One way is that there exists a bounce of the cosmological scale factor in the case of the extra EiBI parameter $\kappa<0$. It is in fact a realization of the bouncing cosmology to avoid the big bang singularity, and it was further studied in e.g~\cite{Avelino:2012ue} later. The other way is that the cosmological scale factor stayed at a minimum size for a long time before inflation in the case of $\kappa>0$. It can be regarded as the emergent universe scenario~\cite{Ellis:2002we, Ellis:2003qz} which can also avoid the big bang singularity.

The emergent universe scenario~\cite{Ellis:2002we, Ellis:2003qz} assumes that the Einstein static universe is the initial state for a past-eternal inflationary cosmological model and then evolves to an inflationary era. In this scenario, there are many attractive advantages. For instance, the horizon problem can be solved before inflation begins. In addition, there is no singularity, no exotic physics is involved, and the quantum gravity regime can even be avoided. Furthermore, it has been proposed that the Einstein static state is favored by  entropy considerations as the initial state for our universe~\cite{Gibbons:1987jt, Gibbons:1988bm, Mulryne:2005ef}.

As is well known, the Einstein static universe is the first cosmological model developed by Einstein~\cite{Einstein:1917ce} in 1917, in which the universe is homogeneous, isotropic, and spatially closed. By adding a positive cosmological constant to Einstein's equations of GR to counteract the attractive effects of gravity on ordinary matter, the universe can be neither expanding nor contracting. However, it was demonstrated by Eddington~\cite{Eddington:1930zz} in 1930 that the Einstein static universe is unstable with respect to homogeneous and isotropic scalar perturbations in GR. In other words, the universe cannot be static, since it must contract or expand in the presence of perturbations. On the other hand, it was also found by Hubble~\cite{Hubble:1929ig} in 1929 that the universe is expanding rather than static, by examining the relation between distance and redshift of galaxies. So, due to Eddington's instability argument and Hubble's astronomical finding, Einstein gave up the idea of static universe (as well as cosmological constant). One can clearly see that the stability analysis played an important role in history.

Recently, the Einstein static universe has been revived to avoid the big bang singularity in the emergent universe scenario. Due to the above historical lesson, the analysis of stability of the Einstein static universe should be thought over. So far, the stability of the Einstein static universe has been studied in vast gravity theories. In GR, this was reconsidered, and it was found that the Einstein static universe can be stable against small inhomogeneous vector and tensor perturbations as well as adiabatic scalar density perturbations if the universe contains a perfect fluid with $w = c_s^2 >1/5$~\cite{Gibbons:1987jt, Gibbons:1988bm, Barrow:2003ni}. Of course, the stability of the Einstein static universe has also been extensively studied in many modified gravities, for example, loop quantum cosmology~\cite{Parisi:2007kv}, $f(R)$ theory~\cite{Boehmer:2007tr, Seahra:2009ft, Goheer:2008tn}, $f(T)$ theory~\cite{Wu:2011xa, Li:2013xea},  modified Gauss-Bonnet gravity~\cite{Bohmer:2009fc, Huang:2015kca}, Brans-Dicke theory~\cite{delCampo:2007mp, delCampo:2009kp, Huang:2014fia, Miao:2016obc}, Horava-Lifshitz theory~\cite{Wu:2009ah, Boehmer:2009yz, Heydarzade:2015hra}, massive gravity~\cite{Parisi:2012cg, Mousavi:2016eof}, braneworld scenario~\cite{Gergely:2001tn, Zhang:2010qwa, Atazadeh:2014xsa}, Einstein-Cartan theory~\cite{Atazadeh:2014ysa}, $f(R, T)$ gravity~\cite{Shabani:2016dhj}, hybrid metric-Palatini gravity~\cite{Boehmer:2013oxa} and so on~\cite{Boehmer:2003iv, Goswami:2008fs, Canonico:2010fd, Zhang:2016obw, Carneiro:2009et, Boehmer:2015ina, Atazadeh:2015zma, Atazadeh:2016yeh}. We refer to e.g.~\cite{Barrow:2003ni} and references therein for more theoretical details of the stability analysis of the Einstein static universe.

Note that it is necessary to consider both the homogeneous and inhomogeneous scalar perturbations in the stability analysis of the Einstein static universe. Although the Einstein static universe is stable against only homogeneous perturbation or only inhomogeneous perturbation in some modified gravities, for instance $f(R)$ theory~\cite{Boehmer:2007tr, Goheer:2008tn, Seahra:2009ft}, it can still be unstable by considering both the homogeneous and inhomogeneous scalar perturbations simultaneously, if the stability conditions for homogeneous and inhomogeneous scalar perturbations do not overlap.

Since the Einstein static universe can exist in EiBI theory as one of the ways to avoid the big bang singularity~\cite{Banados:2010ix}, it is natural to ask ``\,is Einstein static universe stable for perturbations in EiBI theory?\,'' If the corresponding Einstein static universe is unstable, this way to avoid the big bang singularity is not viable in EiBI theory. So, the stability analysis of the Einstein static universe in EiBI theory is interesting. To our best knowledge, this has not been discussed in the literature.

The main aim of the present work is studying the stability of the Einstein static universe filled with perfect fluid in EiBI theory, against both the homogeneous and inhomogeneous scalar perturbations. The rest of this paper is organized as follow. In Sec.~\ref{sec2}, we give a brief review of EiBI theory and its equations of motion. In Sec.~\ref{sec3}, we obtain the exact Einstein static universe solution in the context of EiBI theory with perfect fluid. In Sec.~\ref{sec4}, we give the linearized equations of motion and discuss the stability against homogeneous and inhomogeneous scalar perturbations for spatially closed and open universe, respectively. Finally, we conclude this paper in Sec.~\ref{sec5}.

%============================= section 2 ===================================

\section{EiBI theory}\label{sec2}

Following e.g.~\cite{Banados:2010ix}, here we briefly review EiBI theory. The action of EiBI theory is given by
\be{action}
{\cal S} = \frac{2}{\kappa} \int d^4 x \left(\sqrt{-\det\left(g_{\mu\nu} +\kappa R_{\mu\nu}(\Gamma)\right)} -\lambda \sqrt{-g} \right) +{\cal S}_m \,,
\ee
where ${\cal S}_m$ is the action of matters, $g$ represents the determinant of $g_{\mu\nu}$, $R_{\mu\nu}(\Gamma)$ denotes the symmetric part of the Ricci tensor built with affine connection $\Gamma$, $\lambda$ is a dimensionless constant. Note that we set $8\pi G =1$ throughout this work. $\lambda$ is taken to be $\lambda= \kappa \Lambda + 1$ (where $\kappa$ is a constant with inverse dimension to that of cosmological constant $\Lambda$), so the theory can recover to GR. Actually, in the limit of $\kappa R \ll g$ with ${\cal S}_m = 0$, EiBI theory reduces to GR. On the other hand, in the limit of $\kappa R \gg g$  with ${\cal S}_m = 0$, it reduces to Eddington's theory.

It is worth noting that the metric and affine connection are independent in EiBI theory. By varying $g_{\mu\nu}$ and ${\Gamma^\rho}_{\mu\nu}$ respectively, the equations of motion are given by
\begin{gather}
\frac{\sqrt{-q}}{\sqrt{-g}} \, q^{\mu\nu} -\lambda \, g^{\mu\nu}= -\kappa \, T^{\mu\nu} \,, \label{eom1}  \\[1mm]
\hat{\nabla}_\rho \left(\sqrt{-q} \, q^{\mu\nu} \right)  =0 \,,  \label{eom2}
\end{gather}
where the energy-momentum tensor $T_{\mu\nu} = -\frac{1}{\sqrt{-g}}\frac{\delta {\cal S}_m}{\delta g^{\mu\nu}}$ , $\hat{\nabla}$ is the covariant derivative defined in terms of affine connection ${\Gamma^\rho}_{\mu\nu}$, and $q_{\mu\nu} = g_{\mu\nu} +\kappa R_{\mu\nu}(\Gamma)$ is an auxiliary metric. Note that the indices are raised and lowered by metric $g^{\mu\nu}$ and $g_{\mu\nu}$ unless otherwise stated. $q$ and $q^{\mu\nu}$ denote the determinant and inverse of $q_{\mu\nu}$ respectively, satisfying
\be{qrelation}
q_{\mu\lambda} q^{\lambda\nu} = {\delta_\mu}^\nu \,, \quad q^{\mu\lambda} q_{\lambda\nu} = {\delta^\mu}_\nu \,,
\ee
where ${\delta_\mu}^\nu$ is the Kronecker symbol. Note that we use the Einstein notation throughout this work. Eq.~(\ref{eom2}) is automatically satisfied if the affine connection ${\Gamma^\rho}_{\mu\nu}$ is compatible with the auxiliary metric $q_{\mu\nu}$,
\begin{equation*}
{\Gamma^\rho}_{\mu\nu} = \frac12 q^{\rho\sigma} (q_{\mu\sigma,\nu} +q_{\nu\sigma,\mu} -q_{\mu\nu, \sigma}) \,,
\end{equation*}
where the comma represents partial derivative. So, the auxiliary metric $q_{\mu\nu}$ can be rewritten by
\be{qmunu}
q_{\mu\nu} = g_{\mu\nu} +\kappa R_{\mu\nu}(q)  \,.
\ee
By multiplying $q^{\mu\nu}$, the trace of Eq.~(\ref{qmunu}) is given by
\be{qtrace}
\quad g_{\mu\nu} q^{\mu\nu} +\kappa R(q) = 1 \,.
\ee
We refer to e.g.~\cite{Banados:2010ix,Yang:2017puy,BeltranJimenez:2017doy} and references therein for more details of EiBI theory.

%============================= section 3 ===================================

\section{Einstein static universe in EiBI theory}\label{sec3}

We consider a Friedmann-Robertson-Walker (FRW) metric
\be{FRW}
ds^2 = -dt^2 +a(t)^2 \left[\frac{dr^2}{1-K r^2} +r^2 \left(d\theta^2 +\sin^2\theta d\phi^2\right)\right] \,,
\ee
where $K= +1,\, 0,\, -1$ denote the metric for the closed, flat and open universe, respectively. To obtain the Einstein static universe, we let the scale factor $a(t)=a_0 = \textup{const}.\ne 0$, and then Eq.~(\ref{FRW}) reduces to the metric of the Einstein static universe
\be{esu}
ds^2 = -dt^2 +a_0^2 \left[\frac{dr^2}{1-K r^2} +r^2 \left(d\theta^2 +\sin^2\theta d\phi^2\right)\right] \,.
\ee
The auxiliary metric can be written as
\be{qesu}
ds_q^2 = -X^2 dt^2 +a_0^2 Y^2 \left[\frac{dr^2}{1-K r^2} +r^2 \left(d\theta^2 +\sin^2\theta d\phi^2\right)\right] \,,
\ee
where $X$ and $Y$ are both background quantities. We assume that the matter is perfect fluid, and the corresponding energy-momentum tensor reads
\be{fluid}
T^{\mu\nu} = (\rho_0 + P_0) u^\mu u^\nu +P_0 g^{\mu\nu} \,, \quad \textup{with} \quad P_0 = w \rho_0 \,,
\ee
where $\rho_0$ and $P_0$ represent energy density and pressure respectively, $w$ is the constant equation-of-state parameter, and velocity 4-vector $u^\mu$ is given by
\be{velocity}
u^\mu = \left(1,\, 0,\, 0,\, 0\right) \,, \quad \textup{satisfying} \quad u^\mu u_\mu = -1 \,.
\ee
Substituting Eqs.~(\ref{esu}) -- (\ref{velocity}) into (\ref{eom1}) and (\ref{qmunu}), we obtain
\be{sol}\begin{split}
&X^2 = 1 \,, \quad Y^2 = 1+\frac{2 K \kappa}{a_0^2} \,, \\[1.5mm]
&\rho_0 = \frac{2 K \sqrt{a_0^2 +2 K \kappa}}{a_0^3 (1+w)} \,, \quad \Lambda = \frac{\sqrt{a_0^2 +2 K \kappa} \left[a_0^2(1+w) +2K\kappa w \right] - a_0^3 (1+w)}{a_0^3 \kappa (1+w)} \,.
\end{split}\ee
Note that $a_0^2 +2 K \kappa >0$ is required. %In the limit of $\kappa \rightarrow 0$, it is easy to see that $\rho$ and $\Lambda$ reduce to
The first-order Taylor series expansion of $\rho_0$ and $\Lambda$ around $\kappa = 0$ are given by
\be{limrholam}
\rho_0 = \frac{2 K }{a_0^2 (1+w)} +\frac{2 K^2 \kappa}{a_0^4 (w+1)} + {\cal O}(\kappa^2) \,, \quad \Lambda = \frac{K (1+3 w)}{a_0^2 (1+ w)} +\frac{K^2 (3 w-1) \kappa}{2 a_0^4 (w+1)} + {\cal O}(\kappa^2) \,,
\ee
where the zeroth-order values are the same as the ones in GR~\cite{Barrow:2003ni, Boehmer:2007tr}. Eq.~(\ref{sol}) is the necessary and sufficient condition of the existence of the Einstein static universe in EiBI theory. For given values of $\kappa$, $\rho_0$, $\Lambda$, $w$ and $K$, we can determine the value of $a_0$ from Eq.~(\ref{sol}). In the case of $K=0$ and $\rho_0 \ne 0$, Eq.~(\ref{sol}) cannot be satisfied, and it means that there is no flat Einstein static universe filled with perfect fluid in EiBI theory. So, we only discuss the closed ($K=+1$) and open ($K=-1$) Einstein static universe in the followings.

\vspace{-3mm} % used here just for a more comfortable typesetting

%============================= section 4 ===================================

\section{Stability analysis}\label{sec4}

%============================= section 4.1 ===================================

\subsection{Linearized EiBI theory}\label{sec41}

Now we study the stability of the Einstein static universe in EiBI theory. The background and perturbation components will be denoted by a bar and a tilde, respectively. At first, we try to obtain the linearized equations of motion. The perturbed metric can be written as
\be{pertg}
g_{\mu\nu} = \bar{g}_{\mu\nu} + h_{\mu\nu} \,,
\ee
where $\bar{g}_{\mu\nu}$ is the background metric given by Eq.~(\ref{esu}), and $h_{\mu\nu}$ is a small perturbation. Now the indices are lowered and raised by the background metric unless otherwise stated. By using the relation $g^{\mu\nu} g_{\nu\lambda} ={\delta^\mu}_\lambda$, the inverse metric is perturbed by
\be{invg}
\widetilde{g}^{\mu\nu} = -\bar{g}^{\mu\rho} \bar{g}^{\nu\sigma} h_{\rho\sigma} \,.
\ee
Following Eq.~(\ref{pertg}), the perturbed auxiliary metric $q_{\mu\nu}$ can also be written as
\be{pertq}
q_{\mu\nu} =\bar{q}_{\mu\nu} +\widetilde{q}_{\mu\nu}  \,.
\ee
By using Eq.~(\ref{qrelation}), the inverse auxiliary metric $q^{\mu\nu}$ is given by $q^{\mu\nu} = \bar{q}^{\mu\nu} +\widetilde{q}^{\mu\nu} = \bar{q}^{\mu\nu} -\bar{q}^{\mu\rho} \bar{q}^{\nu\sigma} \widetilde{q}_{\rho\sigma}$, and then we find
\be{pertqinv}
  \widetilde{q}^{\mu\nu} = -\bar{q}^{\mu\rho} \bar{q}^{\nu\sigma} \widetilde{q}_{\rho\sigma}\,.
\ee
The perturbed determinants induced by the perturbation of the metric are given by
\be{}
\sqrt{-g} = \sqrt{-\bar{g}} \left(1+\frac{h}{2} \right) \quad \textup{and} \quad \sqrt{-q} = \sqrt{-\bar{q}} \left(1 + \frac12 \bar{q}^{\mu\nu}\widetilde{q}_{\mu\nu} \right)  \label{pertdet}  \,,
\ee
where $h= h_{\mu\nu} \bar{g}^{\mu\nu}$, and then we have
\be{}
\frac{\sqrt{-q}}{\sqrt{-g}} = \frac{\sqrt{-\bar{q}}}{\sqrt{-\bar{g}}} \left(1 + \frac12 \bar{q}^{\mu\nu}\widetilde{q}_{\mu\nu} -\frac{h}{2} \right)  \label{qoverg} \,.
\ee
 The perturbation of Ricci tensor $\widetilde{R}_{\mu\nu}(q)$ induced by the perturbation of the metric is given by
\be{pertric}
\widetilde{R}_{\mu\nu}(q) = \frac12 \bar{q}^{\rho\sigma} \left( \bar{\nabla}^{(q)}_\rho \bar{\nabla}^{(q)}_{\mu} \widetilde{q}_{\nu\sigma} + \bar{\nabla}^{(q)}_\rho \bar{\nabla}^{(q)}_{\nu} \widetilde{q}_{\mu\sigma} - \bar{\nabla}^{(q)}_\mu \bar{\nabla}^{(q)}_\nu \widetilde{q}_{\rho\sigma} - \bar{\nabla}^{(q)}_{\rho} \bar{\nabla}^{(q)}_{\sigma} \widetilde{q}_{\mu\nu} \right)\,,
\ee
where $\bar{\nabla}^{(q)}$ is the covariant derivative which is compatible with $\bar{q}_{\mu\nu}$. The perturbation of Ricci scalar $\widetilde{R}(q)$ is given by
\be{pertr}
\widetilde{R}(q) = \bar{q}^{\mu\nu} \widetilde{R}_{\mu\nu}(q) +\widetilde{q}^{\mu\nu} \bar{R}_{\mu\nu}(q)  \,.
\ee
For perfect fluid, the perturbations of energy density and pressure are $\widetilde{\rho}$ and $\widetilde{P} = w \widetilde{\rho}$, respectively. The perturbations of the velocity are given by
\be{pertu}
  \widetilde{u}_0 = \widetilde{u}^0 = \frac{h_{00}}{2}   \,, \quad \widetilde{u}^i = \bar{g}^{ij} \, \widetilde{u}_j = \bar{g}^{ij} \bar{\nabla}_j U  \,,
\ee
where the indices ``\,0\," and ``\,$i, j$\," denote the time and space components, respectively. The perturbed energy-momentum tensor is given by
\be{pertT}
\widetilde{T}^{\mu\nu} = P_0 \, \widetilde{g}^{\mu\nu} +\widetilde{P} \, \bar{g}^{\mu\nu} +  (\widetilde{\rho} + \widetilde{P}) u^\mu u^\nu +(\rho_0 + P_0) \widetilde{u}^\mu u^\nu +(\rho_0 + P_0) u^\mu \widetilde{u}^\nu  \,,
\ee
where $u^\mu$ represents the background components and it is given by Eq.~(\ref{velocity}). Considering the above expressions, the linearized equations of Eqs.~(\ref{eom1}), (\ref{qmunu}) and (\ref{qtrace}) read
\begin{gather}
 \frac{\sqrt{-\bar{q}}}{\sqrt{-\bar{g}}} \left( \frac12 \, \bar{q}^{\mu\nu} \bar{q}^{\rho\sigma} \widetilde{q}_{\rho\sigma} -\frac12 \bar{q}^{\mu\nu} h +\widetilde{q}^{\mu\nu}  \right) -\lambda \, \widetilde{g}^{\mu\nu}  = -\kappa \, \widetilde{T}^{\mu\nu} \,, \label{lineom1}\\[1mm]
 \widetilde{q}_{\mu\nu} = \widetilde{g}_{\mu\nu} +\kappa \, \widetilde{R}_{\mu\nu}(q)  \,, \label{lineom2} \\[1.5mm]
 g_{\mu\nu} \, \widetilde{q}^{\mu\nu} + h_{\mu\nu} \, \bar{q}^{\mu\nu} + \kappa \, \widetilde{R}(q) = 0 \,. \label{lintr}
\end{gather}

For our purpose, we consider scalar perturbations in the Newtonian gauge. ${h_\mu}^\nu$ is given by
\be{hdef}
{h_\mu}^\nu = \textup{diag} \left(-2 \Psi ,\, 2 \Phi ,\, 2 \Phi ,\, 2 \Phi\right) \,.
\ee
Note that $\Psi$, $\Phi$, $\widetilde{\rho}$ and $U$ are all functions of $t$, $r$, $\theta$, $\phi$. For scalar perturbations, it is useful to perform a harmonic decomposition~\cite{Harrison:1967zza},
\be{decomposition}
\begin{split}
\Psi = \Psi_n(t)\, Y_n (r, \theta, \phi) \,,\quad \Phi = \Phi_n(t) \,Y_n (r, \theta, \phi) \,,\\[1mm]
\widetilde{\rho} = \rho_0 \, \xi_n(t) \,Y_n (r, \theta, \phi) \,,\quad U = U_n(t) \,Y_n (r, \theta, \phi) \,.
\end{split}
\ee
In these expressions, summations over comoving wave number $n$ are implied. The harmonic function $Y_n (r, \theta, \phi)$ satisfies~\cite{Harrison:1967zza, Huang:2015kca}
\be{laplacian}
\Delta Y_n (r, \theta, \phi) =  -k^2 Y_n (r, \theta, \phi) \,,
%\begin{cases}
%-n(n+2) Y_n (r, \theta, \phi) \quad  K =1\\
%-(n^2+1) Y_n (r, \theta, \phi)  \quad  K =-1
%\end{cases}  \,,
\ee
where $\Delta$ is the Laplacian operator, and $k$ is the separation constant. For the spatially closed universe corresponding to $K=+1$, we have $k^2 =n(n+2)$~\cite{Harrison:1967zza} where the modes are discrete $(n= 0,\, 1,\, 2\dots)$. For the spatially open universe corresponding to $K=-1$, we have $k^2 =n^2+1$~\cite{Harrison:1967zza} where $n\ge 0$. Formally, $n = 0$ gives a spatially homogeneous mode and $n = 1,\, 2\dots$ correspond to spatially inhomogeneous modes for both the closed and open universe~\cite{Barrow:2003ni, Huang:2015kca}.

Substituting Eqs.~(\ref{hdef}) and (\ref{decomposition}) into (\ref{lineom1}) -- (\ref{lintr}), after some algbra, we find
\begin{align}
U_n &=\frac{a_0^2 \Phi_n^\prime(t)}{K} + \frac{\kappa (a_0^2 -a_0^2 w -2 w \kappa K)\, \xi_n^\prime(t)}{2(1+w) (a_0^2+2 \kappa K)} \,, \\[2mm]
\xi_n(t) &= \frac{2 a_0^2 (1+w) (k^2 -3 K) (a_0^2 +2 \kappa K) \Phi_n(t)}{K (2 a_0^4 +a_0^2 \kappa (4 K +k^2 (w-1)) +2 K w \kappa k^2)} \,, \\[2mm]
\begin{split}
\Psi_n(t) &= \frac{2 a_0^4 + 6 K \kappa^2 (4 K - k^2) w + a_0^2 \kappa (4 K - k^2) (1 + 3 w) }{2 a_0^4 + a_0^2 \kappa (4 K + k^2 (w-1)) +2 K \kappa^2 k^2 w} \Phi_n(t) \\[1mm]
&\quad-\frac{2 \kappa (2 a_0^4 + 6 K^2 \kappa^2 w + a_0^2 K \kappa (1 + 3 w)) }{2 a_0^4 + a_0^2 \kappa (4 K + k^2 (w-1)) + 2 K \kappa^2 k^2 w} \Phi_n^{\prime \prime} (t)\,,
\end{split} \label{psin}
\end{align}
where $\Phi_n(t)$  satisfies a second order ordinary differential equation,
\be{ode}
\Phi_n^{\prime\prime}(t) +  Z \Phi_n(t) =0 \,,
\ee
in which a prime represents a derivative with respect to time $t$, and
\be{Zdef}
Z = \frac{k^2 \left(2 a_0^4 w+7 a_0^2 K \kappa w+a_0^2 K \kappa+6 K^2 \kappa^2 w\right)-2 K \left(a_0^2+2 K \kappa \right) \left(3 a_0^2 w+a_0^2+6 K \kappa w\right)}{\left(a_0^2+2 K \kappa\right) \left(2 a_0^4+3 a_0^2 K \kappa w+a_0^2 K \kappa+6 K^2 \kappa^2 w\right)} \,.
\ee
Substituting Eqs.~(\ref{ode}) and (\ref{Zdef}) into~(\ref{psin}), we find
\be{psin2}
\Psi_n(t) = \frac{\Phi_n(t)}{1 +2 K \hat{\kappa}}  \,,
\ee
where we have introduced $\hat{\kappa}\equiv\kappa / a_0^2$ for simplicity. %It is easy to see that $\Psi_n(t)=\Phi_n(t)$ in the cases of $\kappa=0$ or $K=0$.
 To analyze the stability of the Einstein static universe in EiBI theory, we need to discuss the existence condition of the oscillating solution of Eq.~(\ref{ode}). From Eq.~(\ref{ode}), it is easy to see that the stability condition for the Einstein static universe is
\be{Zcond}
Z>0  \,.
\ee
Apart from this condition, the energy density of perfect fluid $\rho_0$ and the scale factor of the Einstein static universe $a_0$ should be real and positive,
\be{rhoacond}
\rho_0>0\,,   \quad  a_0 > 0\,.
\ee
In the following subsections, we discuss both the homogeneous and inhomogeneous perturbations for the closed ($K=+1$) and open ($K=-1$) Einstein static universes, respectively. Note that actually all the equations contain $a_0$ in the forms of $a_0^2>0$ or $a_0^4>0$, so the sign of $a_0$ does not affect the analysis. In the following, we do not mention the condition $a_0>0$ again.

%============================= section 4.2 ===================================

\subsection{$K=+1$}\label{sec42}
In the case of $K=+1$, corresponding to the closed Einstein static universe, the stability needs
\be{cond1}
%\begin{split}
Z = \frac{k^2 \left(2 w+7 \hat{\kappa} w+\hat{\kappa} +6 \hat{\kappa}^2 w\right)-2 \left(1 +2 \hat{\kappa} \right) \left(3 w+ 1+6 \hat{\kappa} w\right)}{a_0^2 \left(1+2 \hat{\kappa}\right) \left(2+3 \hat{\kappa} w+\hat{\kappa}+6 \hat{\kappa}^2 w\right)} >0\,,
\quad \rho_0 = \frac{2 \sqrt{1 +2 \hat{\kappa}}}{a_0^2 (1+w)}>0  \,.% \quad a_0>0\,.
%\end{split}
\ee
Clearly, $\hat{\kappa}>-1/2$ and $w>-1$ are required by $\rho_0>0$. The homogeneous scalar perturbation corresponds to $n=0$, namely $k^2 = 0$. The first inhomogeneous mode ($n = 1$) corresponds to a gauge degree of freedom related to a global rotation, which reflects the freedom to change the four-velocity of fundamental observers~\cite{Barrow:2003ni,Seahra:2009ft, Huang:2015kca}. So, the physical inhomogeneous modes have $n\ge2$ and hence $k^2=n(n+2)\ge 8$. Considering Eq.~(\ref{cond1}), the Einstein static universe could be stable in the following regions:
\begin{itemize}
\item  Case 1: For  $-1/2 < \hat{\kappa} \le -1/3$, the stability requires
	 \begin{flalign}
	&\textup{1.1)} \quad w> -\frac{2 +\hat{\kappa}}{3 \hat{\kappa} \left(1 +2 \hat{\kappa}\right)} \,
	 \quad \textup{with} \quad
	 k^2 =0  \,,& \label{hom1} \\[1mm]
	&\textup{1.2)}  \quad  f(k^2) <w< -\frac{2 +\hat{\kappa}}{3 \hat{\kappa} \left(1+2 \hat{\kappa} \right)} \,
	 \quad \textup{with} \quad
	  k^2 \ge 8  \label{inhom1}  \,,&
	\end{flalign}
	where
	\be{fk}
	f(k^2) \equiv \frac{2 -\hat{\kappa} {k^2}+4 \hat{\kappa} }{\left( 1+2 \hat{\kappa} \right) \left(2 k^2 -6 +3 \hat{\kappa} k^2 -12 \hat{\kappa} \right)} \,.
	\ee
	However, in this case, it is easy to see that the closed Einstein static universe cannot be stable against both the homogeneous ($k^2=0$) and inhomogeneous ($k^2\ge 8$) scalar perturbations simultaneously, since the stability conditions given in Eqs.~(\ref{hom1}) and (\ref{inhom1}) do not overlap.

\item Case 2: For $-1/3 < \hat{\kappa} <0$, the stability requires
	 \begin{flalign}
	 	&\textup{2.1)} \quad -1< w < f(k^2) \,
	 	 \quad \textup{with} \quad
	 	 k^2 =0  \,, &\label{hom2}\\[1.5mm]
	 	&\textup{2.2)} \quad w> -\frac{2 +\hat{\kappa}}{3 \hat{\kappa} \left(1+2 \hat{\kappa}\right)} \,
	 	 \quad \textup{with} \quad
	 	  k^2 =0  \,,& \label{hom3} \\[1mm]
	 	&\textup{2.3)} \quad  f(k^2) <w< -\frac{2 +\hat{\kappa}}{3 \hat{\kappa} \left(1+2 \hat{\kappa} \right)}   \,
	 	\quad \textup{with} \quad
	 	k^2 \ge 8   \label{inhom2}   \,.&
	 \end{flalign}
	 Unfortunately, in this case, we find that
	 \be{ksqcond}
	 f(k^2 \ge 8) -f(k^2=0) = \frac{2}{3(1+2 \hat{\kappa}) (2+3\hat{\kappa} -6(1+2\hat{\kappa})/k^2) }>\frac{2}{3(1+2 \hat{\kappa}) (2+3\hat{\kappa}) } >0
	 \ee
always holds. So, there is no stable region for the closed Einstein static universe against both the homogeneous ($k^2=0$) and inhomogeneous ($k^2\ge 8$) scalar perturbations simultaneously, since the stability conditions given in Eqs.~(\ref{hom2}) or (\ref{hom3}) do not overlap with the one given in Eq.~(\ref{inhom2}).

\item Case 3: For $0 \le  \hat{\kappa} \le \ (\sqrt{13}-1)/6 $, the stability requires
	\begin{flalign}
		&\textup{3.1)} \quad -1< w < f(k^2) \,
		\quad \textup{with} \quad
		k^2 =0  \,, & \label{hom4}\\[1mm]
		&\textup{3.2)} \quad w> f(k^2) \,
		\quad \textup{with} \quad
		k^2 \ge 8  \label{inhom3} \,.&
	\end{flalign}
	When $\hat{\kappa}=0$, EiBI theory reduces to GR, and one can see that the stability conditions (\ref{hom4}) and (\ref{inhom3}) of the closed Einstein static universe in EiBI theory against the homogeneous~($k^2=0$) and inhomogeneous~($k^2\ge 8$) scalar perturbations reduce to $-1<w<-1/3$ and $w > 1/(k^2 - 3)$ respectively, which are the same as the ones in GR~\cite{Barrow:2003ni}.

In the general case of $0 \le  \hat{\kappa} \le \ (\sqrt{13}-1)/6 $, one can easily check that Eq.~(\ref{ksqcond}) is still satisfied, and hence the closed Einstein static universe cannot be stable against both the homogeneous ($k^2=0$) and inhomogeneous ($k^2\ge 8$) scalar perturbations simultaneously, since the stability conditions given in Eqs.~(\ref{hom4}) and (\ref{inhom3}) do not overlap.

\item Case 4: For  $\hat{\kappa} > (\sqrt{13}-1)/{6} $, the stability requires
	\begin{flalign}
		&\textup{4.1)} \quad -\frac{2 +\hat{\kappa}}{3 \hat{\kappa} \left(1+2 \hat{\kappa} \right)} < w < f(k^2)
		\quad \textup{with} \quad
		 k^2 =0  \,, & \label{hom5}\\[1mm]
		&\textup{4.2)} \quad -1 < w < -\frac{2 +\hat{\kappa}}{3 \hat{\kappa} \left(1+2 \hat{\kappa} \right)} \,
		\quad \textup{with} \quad
	    k^2 \ge	8  \,,&  \label{inhom4} \\[1.5mm]
		&\textup{4.3)} \quad w > f(k^2) \,
		\quad \textup{with} \quad
		k^2 \ge 8 \label{inhom5} \,.&	
	\end{flalign}
	Again, Eq.~(\ref{ksqcond}) is still satisfied in this case. So, there is no stable region for the closed Einstein static universe against both the homogeneous ($k^2=0$) and inhomogeneous ($k^2\ge 8$) scalar perturbations simultaneously, since the stability conditions given in Eqs.~(\ref{inhom4}) or (\ref{inhom5}) do not overlap with the one given in Eq.~(\ref{hom5}).
\end{itemize}

\begin{center}
 \begin{figure}[tb]
 \centering
 \vspace{-10mm} % used here just for a more comfortable typesetting
 \includegraphics[width=0.49\textwidth]{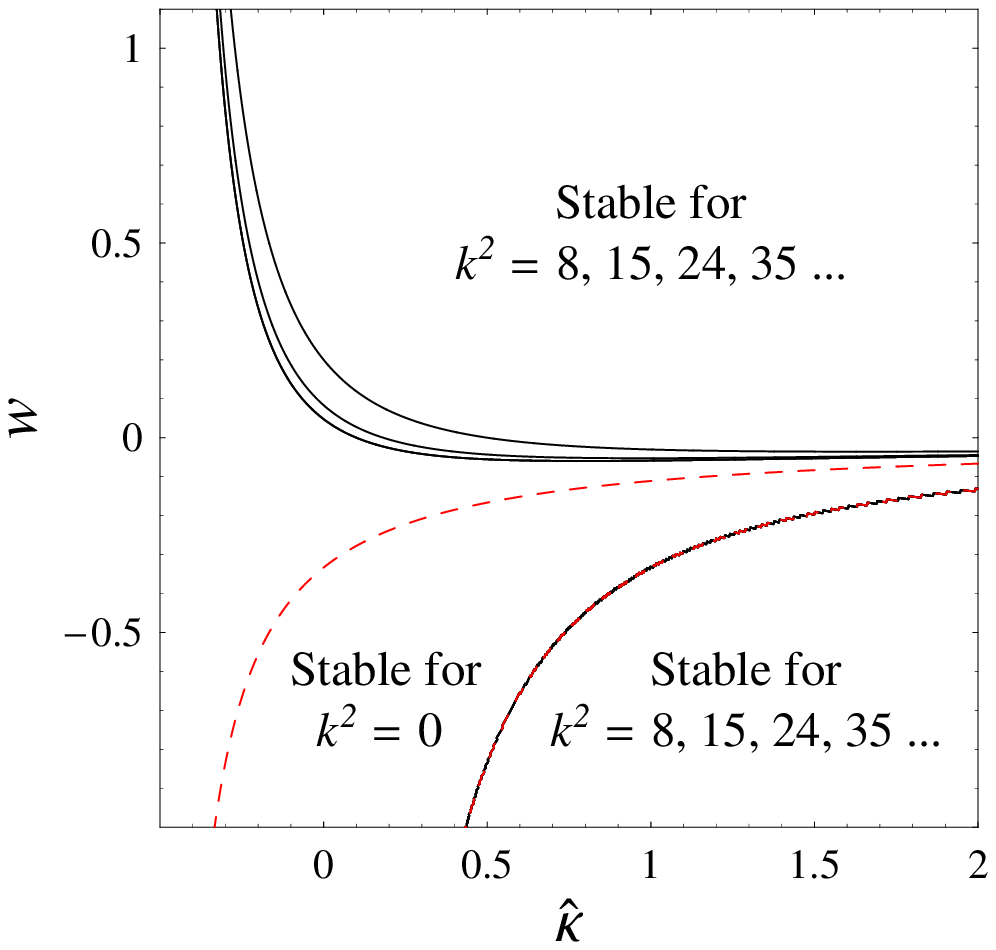}
 \caption{\label{fig1}
 The stable regions with $Z>0$ for the case of the spatially closed universe ($K=+1$). The red dashed lines and the black solid lines are the contours with $Z=0$ for the homogeneous ($k^2=0$) and inhomogeneous ($k^2=n(n+2)\ge 8$) scalar perturbations, respectively. See the text for details.}
 \end{figure}
 \end{center}

Alternatively, we can discuss the stability in general. Note that the quantity $Z$ given in Eq.~(\ref{cond1}) is a function of $\hat{\kappa}$, $w$ and $k^2$. In Fig.~\ref{fig1}, we plot the stable regions with $Z>0$ for the case of the spatially closed universe (we thank the referee for this suggestion). The closed Einstein static universe is stable against the homogeneous scalar perturbation in the region between the red dashed lines with $k^2=0$, while it is stable against the inhomogeneous scalar perturbation in the regions on the right-hand side of the black solid lines with various $k^2=n(n+2)\ge 8$. From Fig.~\ref{fig1}, it is easy to see that these two stable regions for the homogeneous and inhomogeneous scalar perturbations do not overlap unfortunately.

In summary, for all the physical parameters $\hat{\kappa}>-1/2$ and $w>-1$ (required by $\rho_0>0$, nb.~Eq.~(\ref{cond1})), the closed Einstein static universe cannot be stable against both the homogeneous and inhomogeneous scalar perturbations simultaneously. Thus, in the case of the spatially closed universe ($K=+1$), although the closed Einstein static universe solution can exist, but it is unstable unfortunately, and hence the emergent universe scenario is not viable in EiBI theory.

%============================= Fig. 1 =================================

%======================================================================

\vspace{-11mm} % used here just for a more comfortable typesetting

%============================= section 4.3 ===================================

\subsection{$K = -1$}\label{sec43}

For the spatially open universe corresponding to  $K=-1$, the stability conditions reduce to
\be{cond2}
%\begin{split}
	Z = \frac{2 \left(1 -2 \hat{\kappa} \right) \left(3 w+1 -6 \hat{\kappa} w\right)+k^2 \left(2 w-7  \hat{\kappa} w- \hat{\kappa}+6 \hat{\kappa}^2 w\right)}{a_0^2 \left(1-2 \hat{\kappa} \right) \left(2 -3 \hat{\kappa} w-\hat{\kappa} +6 \hat{\kappa}^2 w\right)} >0\,,% \\[2mm]
\quad	\rho_0 = -\frac{2 \sqrt{1 -2 \hat{\kappa}}}{a_0^2 (1+w)}>0  \,.% \quad a_0>0\,.
%\end{split}
\ee
Obviously, $\hat{\kappa}<1/2$ and $w<-1$ are required by $\rho_0>0$. There is no stable region in GR corresponding to $\kappa = 0$ (equivalently $\hat{\kappa}=0$) in EiBI theory, because $Z=(1+w(3+k^2))/a_0^2<0$ for $w<-1$ in this case. However, if $\hat{\kappa} \ne 0$, the open Einstein static universe can be stable in the following regions:
\begin{flalign}
 &\textup{Case 5:} \quad w< -1 \quad  \textup{with} \quad  \hat{\kappa} \le -\frac{\sqrt{13}-1}{6} \quad  \textup{and} \quad  k^2 \ge 1\,; & \label{open1}\\[1mm]
 &\textup{Case 6:} \quad  w< \frac{2 -\hat{\kappa}}{3 \hat{\kappa} \left(1 -2 \hat{\kappa}\right)} \quad  \textup{with} \quad  -\frac{\sqrt{13}-1}{6}  < \hat{\kappa} <0 \quad  \textup{and} \quad  k^2 \ge 1\,.  & \label{open2}
\end{flalign}
In case 6, we have $w < w_{\textup{max}} = -1$. The open Einstein static universe can be stable against both the homogeneous ($k^2=n^2+1=1$) and inhomogeneous ($k^2=n^2+1>1$) scalar perturbations simultaneously, under the conditions~(\ref{open1}) or (\ref{open2}), which require $\hat{\kappa}<0$ and phantom matter ($w<-1$) in common.

\begin{center}
 \begin{figure}[tb]
 \centering
 \vspace{-8mm} % used here just for a more comfortable typesetting
 \includegraphics[width=0.49\textwidth]{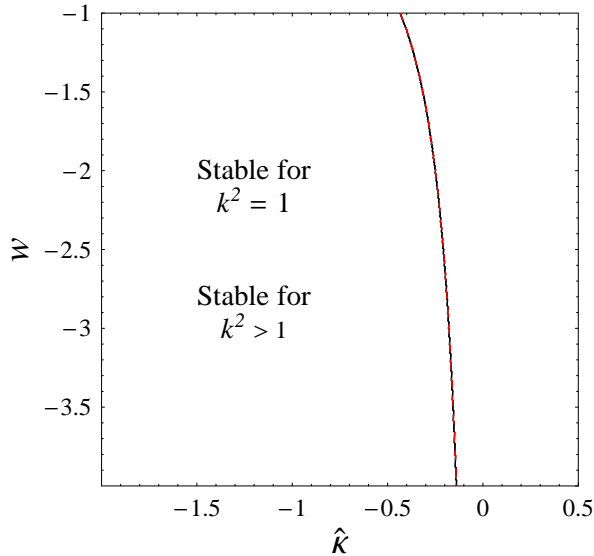}
 \caption{\label{fig2}
 The stable regions with $Z>0$ for the case of the spatially open universe ($K=-1$). The red dashed line and the black solid line are the contours with $Z=0$ for the homogeneous ($k^2=1$) and inhomogeneous ($k^2=n^2+1>1$) scalar perturbations, respectively. See the text for details.}
 \end{figure}
 \end{center}

Similar to Sec.~\ref{sec42}, we can discuss the stability in general. The quantity $Z$ given in Eq.~(\ref{cond2}) is a function of $\hat{\kappa}$, $w$ and $k^2$. In Fig.~\ref{fig2}, we plot the stable regions with $Z>0$ for the case of the spatially open universe (we thank the referee for this suggestion). The open Einstein static universe is stable against the homogeneous ($k^2=1$) and inhomogeneous ($k^2=n^2+1>1$) scalar perturbations in the regions on the left-hand side of the red dashed line and the black solid line, respectively. From Fig.~\ref{fig2}, it is easy to see that these two stable regions for the homogeneous and inhomogeneous scalar perturbations do overlap.

%============================= Fig. 2 =================================

%======================================================================

\vspace{-10mm} % used here just for a more comfortable typesetting

%============================= section 5 ===================================

\section{Conclusion and discussion}\label{sec5}
By considering the realization of the emergent universe scenario in EiBI theory, we study the stability of the Einstein static universe filled with perfect fluid in EiBI theory against both the homogeneous and inhomogeneous scalar perturbations in this work. At first, we obtain the Einstein static universe solution. Then, we derive the linearized equations of motion. Finally, we discuss the stability of the Einstein static universe against both the homogeneous and inhomogeneous scalar perturbations in the spatially closed and open cases, respectively.

We find that there is no spatially flat Einstein static universe filled with perfect fluid in EiBI theory. So, the emergent universe scenario cannot be realized in the flat case. On the other hand, in the spatially closed case, we find that the Einstein static universe cannot be stable against both the homogeneous and inhomogeneous scalar perturbations simultaneously. In other words, the universe cannot be static, since it must contract or expand in the presence of perturbations. So, the emergent universe scenario cannot be viable in the closed case too. As mentioned in Sec.~\ref{sec1}, the big bang singularity could be avoided via the emergent universe scenario or the bouncing scenario in EiBI theory~\cite{Banados:2010ix}. Our results show that in both the spatially flat and closed cases, the emergent universe scenario is no longer viable, but the bouncing scenario still survives so far.

However, the emergent universe scenario in EiBI theory might be viable in the spatially open case. We find that the open Einstein static universe can be stable against both the homogeneous and inhomogeneous scalar perturbations simultaneously, while $\kappa<0$ and $w<-1$ are necessary. As is well known, phantom matter with $w<-1$ violates all the energy conditions, and brings about some undesirable properties. For instance, the universe dominated by phantom will end in a big rip~\cite{Caldwell:2003vq} in the far future. But this is not a problem to us, since in this work we only consider the very early era before inflation. On the other hand, as is also well known, phantom is unstable in the quantum level (see e.g.~\cite{Carroll:2003st,Doran:2002bc}). This might be troublesome in most cosmological models. Fortunately, this is also not a problem to us. Because the Einstein static universe is the initial state in the emergent universe scenario, and if the scale factor of the Einstein static universe $a_0$ is not so small, the quantum effects can be ignored. So, one can safely deal with phantom in the classical level.

In fact, the constant equation-of-state parameter of dark energy as a perfect fluid has been determined to be $w=-1.006\pm 0.045$~\cite{Ade:2015xua} by using the latest observational data of type Ia supernovae (SNIa) and cosmic microwave background (CMB). Thus, $w<-1$ is slightly favored by the cosmological observations. On the other hand, the spatial curvature was determined to be $\Omega_K=0.000\pm 0.005$~\cite{Ade:2015xua} at 95\% confidence level. This means that although a spatially flat universe ($K=0$) is favored, a spatially open universe ($K=-1$) is still consistent with the observational data. So, our results can find slight support from the cosmological observations.

It is worth noting that in principle all types of perturbations should be taken into account when we study the stability of the Einstein static universe. In the present work, we have only considered the scalar perturbations. It is also important to study the stability against tensor and vector perturbations, and we leave this to future work.

%============================= acknowledgements ===================================

\section*{ACKNOWLEDGEMENTS}
We thank the anonymous referee for quite useful comments and suggestions, which helped us to improve this work. S.L.L. is grateful to Prof.~H. L\"u, Prof. Shuang-Qing~Wu, and Dr.~He~Huang, for kind and useful discussions. We also thank Xiao-Bo~Zou, Hong-Yu~Li, Dong-Ze~Xue, Hua-Kai~Deng and Zhao-Yu~Yin for helpful discussions. This work was supported in part by NSFC under Grants No.~11575022 and No.~11175016.

\renewcommand{\baselinestretch}{1.0}

%============================= references ==================================

\end{document}